\documentclass[pre,twocolumn,aps,showpacs]{revtex4}
\newcommand{\bec}[1]{\mbox{\boldmath $ #1$}}
\usepackage{graphicx}
\begin{document}
\title{Effect of heat flux on differential rotation in turbulent
convection}

\author{Nathan Kleeorin}
\email{nat@menix.bgu.ac.il}
\author{Igor Rogachevskii}
\email{gary@menix.bgu.ac.il}
\homepage{http://www.bgu.ac.il/~gary}
\affiliation{Department of Mechanical Engineering, Ben-Gurion
University of the Negev, Beer-Sheva 84105, P. O. Box 653, Israel}
\date{\today}
\begin{abstract}
We studied the effect of the turbulent heat flux on the Reynolds
stresses in a rotating turbulent convection. To this end we solved a
coupled system of dynamical equations which includes the equations
for the Reynolds stresses, the entropy fluctuations and the
turbulent heat flux. We used a spectral $\tau$ approximation in
order to close the system of dynamical equations. We found that the
ratio of the contributions to the Reynolds stresses caused by the
turbulent heat flux and the anisotropic eddy viscosity is of the
order of $\sim 10 \, (L_\rho / l_0)^2$, where $l_{0}$ is the maximum
scale of turbulent motions and $L_\rho$ is the fluid density
variation scale. This effect is crucial for the formation of the
differential rotation and should be taken into account in the
theories of the differential rotation of the Sun, stars and planets.
In particular, we demonstrated that this effect may cause the
differential rotation which is comparable with the typical solar
differential rotation.
\end{abstract}


\maketitle

\section{Introduction}

Solar and stellar magnetic fields are believed to originate in a
dynamo, driven by the joint action of the mean hydrodynamic helicity
of turbulent convection and differential rotation (see, e.g.,
\cite{M78,P79,KR80,ZRS83,BSR05}, and references therein). It was
suggested in \cite{K63} that the differential rotation of the Sun is
caused by an anisotropic eddy viscosity which was described
phenomenologically in \cite{K63,L41,D85,RUD89}. Theory of the
differential rotation based on the idea of the anisotropic eddy
viscosity was developed in a number of papers (see, e.g.,
\cite{KRUD93,D93,KRUD05}, and references therein). However, there is
an additional effect which can strongly modify the differential
rotation. In particular, the direct effect of the turbulent heat
flux on the Reynolds stresses in  a rotating turbulent convection is
crucial for formation of the differential rotation.

The effect of rotation on a hydrodynamic turbulence was studied in
numerous papers (see, e.g., \cite{DS79,KRUD93}). However, a relation
to the turbulent convection was made in previous theories of the
differential rotation only phenomenologically, using the equation
\begin{eqnarray}
\langle {\bf u'}^2 \rangle \propto g \tau_0 \langle u'_z s' \rangle
\;, \label{A6}
\end{eqnarray}
which follows from the mixing-length theory. Here $\langle u'_z s'
\rangle$ is the vertical turbulent heat flux, ${\bf u}'$ and $s'$
are fluctuations of fluid velocity and entropy, ${\bf g}$ is the
acceleration of gravity and $\tau_0$ is the characteristic
correlation time of turbulent velocity field. Equation~(\ref{A6})
implies that the vertical turbulent heat flux plays a role of a
stirring force for the turbulence. However, a more sophisticated
approach implies a solution of a coupled system of dynamical
equations which includes the equations for the Reynolds stresses
$\langle u'_i u'_j \rangle$, the turbulent heat flux $\langle s' \,
u'_i \rangle$ and the entropy fluctuations $\langle s'^2 \rangle$ in
a rotating turbulent convection. The latter has not been taken into
account in the previous theories of the differential rotation.

The goal of this study is to analyze the effect of the turbulent
heat flux on the Reynolds stresses in a rotating turbulent
convection and on formation of the differential rotation. We
demonstrated that this effect is crucial for the formation of the
differential rotation, and it should be taken into account in
theories of the differential rotation of the Sun, stars and planets.
In particular, we found that the ratio of the contributions to the
Reynolds stresses caused by the turbulent heat flux and the
anisotropic eddy viscosity is of the order of $ \sim 10 (L_\rho /
l_0)^2$ for $\Omega \, \tau_0 \leq 1$, where $l_{0}$ is the maximum
scale of turbulent motions (the integral scale of turbulence),
$\Omega$ is the rotation rate, $L_\rho$ is the fluid density
variation scale, i.e., $(\bec{\nabla} \rho_0) / \rho_0 = -
L_\rho^{-1} \, {\bf e}$, $\, \, \rho_0$ is the fluid density and
${\bf e}$ is the unit vector in the direction of the fluid density
inhomogeneity. The turbulent heat flux contribution to the Reynolds
stresses changes its sign when the direction of the vertical
turbulent heat flux changes. This is the key difference from
previous theories of the differential rotation. The effect of the
turbulent heat flux on the Reynolds stresses in a turbulent
convection may cause the differential rotation which is comparable
with the typical differential rotation of the Sun. The data of the
solar differential rotation is obtained from surface observations of
the solar angular velocity (see, e.g., \cite{HH70,SH84}) and from
helioseismology based on measurements of the frequency of $p$-mode
oscillations (see, e.g., \cite{DH86,DG89,T90,SAB98}).

The mechanism of the differential rotation that is associated with
the effect of the turbulent heat flux on Reynolds stresses in a
rotating turbulent convection is as follows. Let us split the total
rotation of fluid into a constant component $\Omega$ (uniform
rotation) and the differential rotation $\delta \Omega$. The uniform
rotation causes the counter-rotation turbulent heat flux (i.e., the
toroidal turbulent heat flux, $\langle s' \, u'_\varphi \rangle $,
directed opposite to the background rotation $\Omega$). Therefore,
there is a correlation of fluctuations of the entropy $s'$ and the
toroidal component of the velocity, $u'_\varphi$. Here $r, \theta,
\varphi$ are the spherical coordinates.

The counter-rotation  turbulent heat flux is similar to the
counter-wind turbulent heat flux (in the direction opposite to the
mean wind) which is well-known in the atmospheric physics. The
counter-rotation  turbulent heat flux arises by the following
reason. In turbulent convection an ascending fluid element has
larger temperature then that of surrounding fluid and smaller
toroidal fluid velocity, while a descending fluid element has
smaller temperature and larger toroidal fluid velocity. This causes
the turbulent heat flux in the direction opposite to the toroidal
mean fluid flow (i.e., opposite to rotation). The counter-rotation
turbulent heat flux is determined by Eq.~(\ref{RB30}) in Section
III.

The entropy fluctuations cause fluctuations of the buoyancy force,
and this results in increase fluctuations of the vertical and
meridional components of the velocity which are correlated with the
fluctuations of the toroidal component of the velocity. These
produce the off-diagonal components of the Reynolds stress tensor,
$\langle u'_r u'_\varphi \rangle $ and $\langle u'_\theta u'_\varphi
\rangle $, and create the toroidal component of the effective force
which causes formation of the differential rotation $\delta \Omega$
in turbulent convection.

This paper is organized as follows. In Section II we formulated the
governing equations, the assumptions, the procedure of the
derivation, and described the effect of the turbulent heat flux on
the Reynolds stresses. In Section III we developed the theory of the
differential rotation based on this effect. In Section IV we made
estimates for the solar differential rotation. In Appendixes A and B
we performed a detailed derivation of the effect of the turbulent
heat flux on the Reynolds stresses in the rotating turbulent
convection.

\section{Effect of the turbulent heat flux on the Reynolds stresses}

In order to study the effect of the turbulent heat flux on the
Reynolds stresses we considered turbulent convection with large
Rayleigh and Reynolds numbers. We employed a mean field approach
whereby the velocity, pressure and entropy are separated into the
mean and fluctuating parts, where the fluctuating parts have zero
mean values. The large-scale fluid motions are determined by the
mean-field equations, which follow from the momentum and entropy
equations for instantaneous fields by averaging over an ensemble of
fluctuations. The mean-field equations are given by
\begin{eqnarray}
\biggl({\partial  \over \partial t} &+& {\bf U} \cdot
\bec{\nabla}\biggr) \, U_{i} = - {\nabla}_{i} \, \biggl({P \over
\rho_0}\biggr) - {\bf g} \, S + 2 \, {\bf U} \times {\bf \Omega}
\nonumber \\
&-& ({\nabla}_{j} + L_\rho^{-1} \, e_{j}) \, \langle u'_{i} \,
u'_{j} \rangle + {\bf f}_{\nu}({\bf U}) \;,
\label{A3} \\
\biggl({\partial  \over \partial t} &+& {\bf U} \cdot \bec{\nabla}
\biggr) \, S =  - ({\nabla}_{j} + L_\rho^{-1} \, e_{j}) \, \langle s
\, u_{i} \rangle
\nonumber \\
&-& {1 \over T_{0}} \, \bec{\nabla} \cdot {\bf F}_{\kappa}({\bf U},
S) \;, \label{A4}
\end{eqnarray}
where Eq.~(\ref{A3}) is written in the reference frame uniformly
rotating with the angular velocity ${\bf \Omega}$. Here the mean
fields ${\bf U}$, $\, P$, $\, T$ and $S$ are the fluid velocity,
pressure, temperature and entropy, respectively, $\rho_0 \, {\bf
f}_{\nu}({\bf U}) $ is the mean molecular viscous force, ${\bf
F}_{\kappa}({\bf U}, S) $ is the mean heat flux that is associated
with the molecular thermal conductivity. The mean fluid velocity
${\bf U}$ for a low Mach number flows satisfies the equation ${\rm
div} \, (\rho_0 \, {\bf U}) = 0$. Equations~(\ref{A3})
and~(\ref{A4}) are written in the anelastic approximation, which is
a combination of the Boussinesq approximation and the condition
${\rm div} \, (\rho_0 \, {\bf U}) = 0 $. The variables with the
subscript $ "0" $ correspond to the hydrostatic nearly isentropic
basic reference state, i.e., $\bec{\nabla} P_{0} = \rho_{0} {\bf g}$
and ${\bf g} \cdot [(\gamma P_{0})^{-1} \bec{\nabla} P_{0} -
\rho_{0}^{-1} \bec{\nabla} \rho_{0}] \approx 0$, where $\gamma$ is
the ratio of specific heats. The turbulent convection is regarded as
a small deviation from a well-mixed adiabatic reference state.

In order to get a closed system of the mean-field equations we have
to determine the dependencies of the Reynolds stresses $\langle
u'_{i}(t,{\bf x}) \, u'_{j}(t,{\bf x}) \rangle $ and the turbulent
heat flux $\langle s'(t,{\bf x}) \, u'_{i}(t,{\bf x}) \rangle$ on
the mean fields. To this end we used equations for fluctuations of
velocity and entropy in a rotating turbulent convection, which are
obtained by subtracting Eqs.~(\ref{A3}) and~(\ref{A4}) for the mean
fields from the corresponding equations for the instantaneous
fields. The equations for fluctuations of velocity and entropy are
given by
\begin{eqnarray}
{\partial {\bf u}' \over \partial t} &=& - ({\bf U} \cdot
\bec{\nabla}) {\bf u}' - ({\bf u}' \cdot \bec{\nabla}) {\bf U} -
\bec{\nabla} \biggl({p' \over \rho_{0}}\biggr)
\nonumber\\
& & - {\bf g} \, s' + 2 {\bf u}' \times {\bf \Omega} + {\bf U}^{N}
\;,
\label{A1} \\
{\partial s' \over \partial t} &=& - {\Omega_{b}^{2} \over g} ({\bf
u}' \cdot {\bf e}) - ({\bf U} \cdot \bec{\nabla}) s' + S^{N} \;,
\label{A2}
\end{eqnarray}
where $ U^{N} = \langle ({\bf u}' \cdot \bec{\nabla}) {\bf u}'
\rangle - ({\bf u}' \cdot \bec{\nabla}) {\bf u}' + {\bf
f}_{\nu}({\bf u}') $ and $ S^{N} = \langle ({\bf u} \cdot
\bec{\nabla}) s \rangle - ({\bf u} \cdot \bec{\nabla}) s - (1/
T_{0}) \, \bec{\nabla} \cdot {\bf F}_{\kappa}({\bf u}', s') $ are
the nonlinear terms which include the molecular dissipative terms,
$\Omega_{b}^{2} = - {\bf g} \cdot \bec{\nabla} S $ is the
Brunt-V\"{a}is\"{a}l\"{a} frequency, $p'$ are fluctuations of fluid
pressure and the fluid velocity fluctuations ${\bf u}'$ satisfy the
equation ${\rm div} \, (\rho_0 \, {\bf u}') = 0$.

To study the rotating turbulent convection we performed the
derivations which include the following steps: (i) using new
variables for fluctuations of velocity ${\bf v} = \sqrt{\rho_0} \,
{\bf u}' $ and entropy $s = \sqrt{\rho_0} \, s'$; (ii) derivation of
the equations for the second moments $M^{(II)}$ of the velocity
fluctuations $\langle v_i \, v_j \rangle$, the entropy fluctuations
$\langle s^2 \rangle$ and the turbulent heat flux $\langle v_i \, s
\rangle$ in the ${\bf k}$ space; (iii) application of the spectral
closure (see Eq.~(\ref{A5}) below) and solution of the derived
second-moment equations in the ${\bf k}$ space; (iv) returning to
the physical space to obtain formulae for the Reynolds stresses and
the turbulent heat fluxes as the functions of the rotation rate
$\Omega$ (see for details, Appendix A).

The second-moment equations include the first-order spatial
differential operators $\hat{\cal N}$  applied to the third-order
moments $M^{(III)}$. A problem arises how to close the system, i.e.,
how to express the set of the third-order terms $\hat{\cal N}
M^{(III)}$ through the lower moments $M^{(II)}$ (see, e.g.,
\cite{O70,MY75,Mc90}). Various approximate methods have been
proposed in order to solve it. A widely used spectral $\tau$
approximation (\cite{O70,PFL76,KRR90,RK04,BK04,BS05}) postulates
that the deviations of the third-moment terms, $\hat{\cal N}
M^{(III)}({\bf k})$, from the contributions to these terms afforded
by the background turbulent convection, $\hat{\cal N}
M_0^{(III)}({\bf k})$, are expressed through the similar deviations
of the second moments, $M^{(II)}({\bf k}) - M_0^{(II)}({\bf k})$:
\begin{eqnarray}
\hat{\cal N} M^{(III)}({\bf k}) &-& \hat{\cal N} M_0^{(III)}({\bf
k})
\nonumber\\
&=& - {M^{(II)}({\bf k}) - M_0^{(II)}({\bf k}) \over \tau(k)} \;,
\label{A5}
\end{eqnarray}
where $\tau(k)$ is the characteristic relaxation time, which can be
identified with the correlation time of the turbulent velocity
field. The background turbulent convection (which corresponds to a
nonrotating and shear-free turbulent fluid flow), is determined by
the budget equations and the general structure of the moments is
obtained by symmetry reasoning. The above procedure (see Appendix A)
yields formulae for the Reynolds stresses and the turbulent heat
flux in the rotating turbulent convection. In particular, this
allowed us to determine the effect of the turbulent heat flux on the
Reynolds stresses.

The differential rotation in the axisymmetric fluid flow is
determined by linearized Eq.~(\ref{A3}) for the toroidal component
$U_\varphi(r, \theta) \equiv r \, \sin \theta \, \delta \Omega$ of
the mean velocity:
\begin{eqnarray}
\rho_0 \, {\partial U_\varphi \over \partial t} &=& {1 \over r^3}
{\partial \over \partial r} (r^3 \sigma_{r \varphi}) + {1 \over r
\sin^2 \theta} {\partial \over \partial \theta} (\sin^2 \theta \,
\sigma_{\theta \varphi})
\nonumber \\
&& + 2 \, \rho_0 \, ({\bf U} {\bf \times} {\bf \Omega})_\varphi \;,
\label{C1}
\end{eqnarray}
where the tensor $\sigma_{ij} = - \langle v_i \, v_j \rangle$ is
determined by the Reynolds stress tensor. In particular,
\begin{eqnarray}
\sigma_{r \varphi} &\equiv& - e_j^\varphi \, e_i^r \, \langle v_i \,
v_j \rangle = \rho_0 \, \nu_{_{T}} \, r \, {\partial \over
\partial r} \biggl({U_\varphi \over r}\biggr)
\nonumber \\
&& + \sigma_{r \varphi}^F + \sigma_{r \varphi}^u \;,
\label{C2}\\
\sigma_{\theta \varphi} &\equiv& - e_j^\varphi \, e_i^\theta \,
\langle v_i \, v_j \rangle = \rho_0 \, \nu_{_{T}} \, {\sin \theta
\over r} \, {\partial \over \partial \theta} \biggl({U_\varphi \over
\sin \theta} \biggr)
\nonumber \\
&& + \sigma_{\theta \varphi}^F + \sigma_{\theta \varphi}^u \;,
\label{C3}
\end{eqnarray}
where ${\bf e}^r$,  $\, {\bf e}^\theta$ and $\, {\bf e}^\varphi$ are
the unit vectors along the radial, meridional and toroidal
directions of the spherical coordinates $r, \theta, \varphi$. There
are three contributions to the tensor $\sigma_{ij} = - \langle v_i
\, v_j \rangle$. In particular, the first term in the right hand
side of Eqs.~(\ref{C2}) and ~(\ref{C3}) describes the isotropic
turbulent viscosity $\propto \nu_{_{T}}$, the second term in
Eqs.~(\ref{C2}) and ~(\ref{C3}) determines the contribution
$\bec{\sigma}^F$ to the Reynolds stresses caused by the turbulent
heat flux, and the third term in Eqs.~(\ref{C2}) and ~(\ref{C3})
determines the contribution $\bec{\sigma}^u$ to the Reynolds
stresses caused by the anisotropy of turbulence due to the
nonuniform fluid density and uniform rotation.

In Eq.~(\ref{C1}) we neglected small molecular viscosity term and we
took into account that in the axisymmetric fluid flow $\partial
({\bf U}, S) /\partial \varphi  = 0$ and $\delta \Omega / \Omega$ is
a small parameter. We assumed that the toroidal component of the
mean velocity is much larger than the poloidal component. This is
typical for the solar and stellar convective zones. We also took
into account that the fluid density is nonuniform in the radial
direction. The first two terms in the right hand side of
Eq.~(\ref{C1}) are the $\varphi$ component of the divergence of the
tensor $\sigma_{ij}$ written for the axisymmetric fluid flow in
spherical coordinates. This is a standard form of the $\varphi$
component of the divergence of a tensor in spherical coordinates
(see, e.g., \cite{BSL02}).

Let us discuss the contribution, $\bec{\sigma}^F$, to the Reynolds
stress tensor caused by the turbulent heat flux. The components of
this tensor, $\sigma_{r \varphi}^F \equiv - e_j^\varphi \, e_i^r \,
\langle v_i \, v_j \rangle^{F}$ and $\sigma_{\theta \varphi}^F = -
e_j^\varphi \, e_i^\theta \, \langle v_i \, v_j \rangle^{F}$, were
determined in Appendix A. They are given by
\begin{eqnarray}
\sigma_{r \varphi}^F &=& {1 \over 6} \, \rho_0 \, \tau_0^2 \, g \,
F_\ast \, \Omega \, \sin \theta \, [\Phi_1(\omega) + \cos^2 \theta
\, \Phi_2(\omega)] \;,
\nonumber\\
\label{B24}\\
\sigma_{\theta \varphi }^F &=& {1 \over 3} \, \rho_0 \, \tau_0^2
\, g \,   F_\ast \, \Omega \, \sin^2 \theta \, \cos \theta \,
\Phi_2(\omega) \;,
\label{B25}
\end{eqnarray}
where the tensor $\langle v_i \, v_j \rangle^{F}$ determines the
contribution to the Reynolds stresses which vanishes when $\langle
s' \, u'_z \rangle \to 0$ (see Eq.~(\ref{B23}) in Appendix A). Here
$F_\ast = \langle s' \, u'_z \rangle$ is the vertical turbulent heat
flux, the parameter $\omega = 8 \, \Omega \, \tau_0 $, $\, \tau_0 =
l_{0} / u_{0}$ is the characteristic correlation time of turbulent
velocity field, and $u_{0}$ is the characteristic turbulent velocity
in the maximum scale of turbulent motions $l_{0}$. The functions
$\Phi_1(\omega)$ and $\Phi_2(\omega)$ are shown in Fig.~1. The
formulae for the functions $\Phi_1(\omega)$ and $\Phi_2(\omega)$ are
given by Eqs.~(\ref{B31}) and~(\ref{B32}) in Appendix A. The
asymptotic formulas for $\sigma_{r \varphi}^F$ and $\sigma_{\theta
\varphi}^F$ for a slow rotation, $8 \, \tau_0 \, \Omega \ll 1$, are
given by
\begin{eqnarray*}
\sigma_{r \varphi}^F &\approx& \rho_0 \, \tau_0^2 \, g \, F_\ast
\, \Omega \, \sin \theta \;,
\\
\sigma_{\theta \varphi}^F &\approx& 4 \, \rho_0 \, \tau_0^4 \, g
\,   F_\ast \, \Omega^3 \, \sin^2 \theta \, \cos \theta \;,
\end{eqnarray*}
and for $8 \, \tau_0 \, \Omega \gg 1$ they are given by
\begin{eqnarray*}
\sigma_{r \varphi}^F &\approx& {1 \over 4 \, \Omega} \, \rho_0 \,
g \, F_\ast  \, \sin^3 \theta \;,
\\
\sigma_{\theta \varphi}^F &\approx& {1 \over 2 \, \Omega} \,
\rho_0 \, g \,   F_\ast \, \sin^2 \theta \, \cos \theta \; .
\end{eqnarray*}

\begin{figure}
\vspace*{2mm} \centering
\includegraphics[width=8cm]{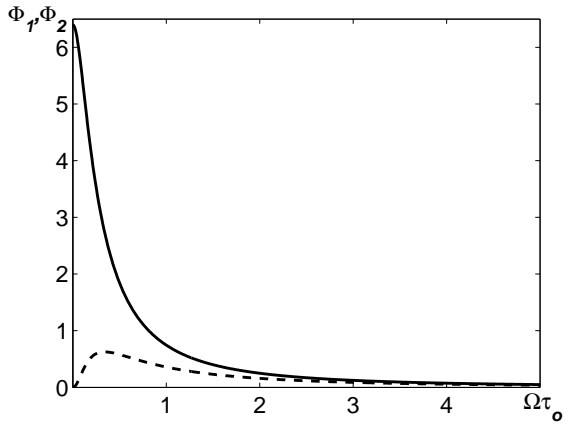}
\caption{\label{Fig1} The functions $\Phi_1(\Omega \, \tau_0)$
(solid) and $\Phi_2(\Omega \, \tau_0)$ (dashed).}
\end{figure}

The contribution to the Reynolds stresses, $\bec{\sigma}^u$, in
Eqs.~(\ref{C2}) and~(\ref{C3}) is caused by the anisotropy of
turbulence due to the inhomogeneous fluid density and uniform
rotation. The tensor $\bec{\sigma}^u$ determines the anisotropic
eddy viscosity tensor with the nonzero off-diagonal components which
are given by
\begin{eqnarray}
\sigma_{r \varphi}^u &\approx& {l_0^2 \over 15 L_\rho^2} \, (\rho_0
\, u_0^2) \, \,  (\Omega \, \tau_0) \, \, \sin \theta \,
[a_1(\Omega) - a_2(\Omega) \, \cos^2 \theta] \;,
\nonumber \\
\label{A9}\\
\sigma_{\theta \varphi}^u &\approx& {l_0^2 \over 15 L_\rho^2} \,
(\rho_0 \, u_0^2) \, \, (\Omega \, \tau_0) \, \, a_2(\Omega) \,
\sin^2 \theta \, \cos \theta \;,
\label{A10}
\end{eqnarray}
(see, e.g., \cite{KRUD93}), where $a_1(\Omega) \sim - 1$ and
$a_2(\Omega) \sim 2 (\Omega \, \tau_0)^2$ for $\Omega \, \tau_0 \ll
1$, and $a_1(\Omega) \sim O[(\Omega \, \tau_0)^{-3}]$ and
$a_2(\Omega) \sim - (\Omega \, \tau_0)^{-1}$ for $\Omega \, \tau_0
\gg 1$. The contribution to the Reynolds stresses, $\bec{\sigma}^u$,
due to the anisotropic eddy viscosity is smaller than that of
$\bec{\sigma}^F$ due to the turbulent heat flux. In particular, the
ratio $|{\sigma}^F / {\sigma}^u| \sim 10 \, (L_\rho / l_0)^2$ for
$\Omega \, \tau_0 \leq 1$. Note that for fast rotation rates
$(\Omega \, \tau_0 \gg 1)$ the validity of Eqs.~(\ref{A9})
and~(\ref{A10}) is questionable, because the quasi-linear
approximation used in \cite{KRUD93} does not valid for $\Omega \,
\tau_0 \gg 1$.

For large Rayleigh numbers the contributions of anisotropies
decrease in small scales. On the other hand, the rotation introduces
an anisotropy in the turbulent convection. This causes non-zero
off-diagonal components of the Reynolds stress tensor (see
Eqs.~(\ref{B24})-(\ref{A10})) and results in the redistribution of
the turbulent heat flux  on the surface of the rotating body (see
Eq.~(\ref{B30})). Note that the main contribution to the tensor
$\sigma_{ij}$ is at the maximum scale of turbulent motions.
Therefore, the contributions to the tensor $\sigma_{ij}$ which
depend on the Reynolds number are negligibly small. In the present
study we investigated the large-scale effects (the differential
rotation), and the influence of the molecular viscosity and
molecular thermal diffusivity on the large-scale dynamics is very
small in comparison with that of the eddy viscosity and turbulent
thermal diffusivity. Therefore, the contributions to the tensor
$\sigma_{ij}$ which depend on the molecular Prandtl number are also
negligibly small.

\section{Differential rotation}

In the present study for simplicity we have taken into account only
the effect of $\bec{\sigma}^F$ on the differential rotation. Let us
neglect the toroidal component of the Coriolis force in
Eq.~(\ref{C1}). This is valid when the poloidal components of the
mean fluid velocity $U_\theta$ and $U_r$ are much smaller than $ u_0
\, l_0 / L_\rho .$ This condition implies that the last term in the
right hand side of Eq.~(\ref{C1}) is much smaller than other terms.
Here we took into account that $\sigma_{r \varphi}^F \propto \rho_0$
and the fluid density stratification length $L_\rho$ is much smaller
than the solar radius $R_\odot$. Therefore, we neglected the effect
of the meridional circulations on the differential rotation, which
was studied in \cite{KRUD05}, among the others. For simplicity we
also did not take into account the dependence of the turbulent
viscosity on the rate of rotation. After these simplifications,
Eq.~(\ref{C1}) in dimensionless form reads
\begin{eqnarray}
\biggl[{\partial \over \partial t} + {\hat {\cal M}(X) \over r^2} -
\hat {\cal W}(r) \biggr] {\delta \Omega \over \Omega} = {{\cal I}
\over r^2} \;, \label{C4}
\end{eqnarray}
where
\begin{eqnarray*}
\hat {\cal M}(X) &=& (X^2 - 1) {\partial^2 \over \partial X^2} + 4 X
{\partial \over \partial X} \;,
\\
\hat {\cal W}(r) &=& {1 \over \rho_0(r) \,r^4} \, {\partial \over
\partial r} \biggl( \rho_0(r) \,r^4 \, {\partial \over
\partial r} \biggr) \;,
\\
{\cal I}(r) &=& I_0 \, [\, (3 - r \, \Lambda) \, \Phi_1(\omega)  - 2
\, \Phi_2(\omega)
\nonumber\\
&& + \, \, (13 - r \, \Lambda) \, \Phi_2(\omega) \, X^2 \, ] \;,
\end{eqnarray*}
$I_0 = \tau_0^2 \, g \, F_\ast / 6 \, \nu_{_{T}}$, $\, \Lambda =
R_\odot / L_\rho$ and $X = \cos\theta$. Here length is measured in
units of the solar radius $R_\odot$ and time  is measured in units
of $R_\odot^{2} / \nu_{_{T}}$ based on the solar radius and the
turbulent viscosity $\nu_{_{T}}$.

Solution of Eq.~(\ref{C4}) we seek for in the form:
\begin{eqnarray}
{\delta \Omega \over \Omega} = \tilde A + {1 \over \sqrt{\rho_0}}
\sum_{n=0}^{\infty} \, C_{2n}^{3/2}(X) \sum_{m=3}^{\infty} \,
V_{m,2n}(r) \, Q_{m,2n}(t) \;,
\nonumber\\
\label{C5}
\end{eqnarray}
where the function $C_n^{3/2}(X)$ satisfies the equation for the
ultra-spherical polynomials:
\begin{eqnarray}
[\hat {\cal M}(X) - n(n+3)] \, C_n^{3/2}(X) = 0  \; . \label{C6}
\end{eqnarray}
The functions $V_{m,n}(r)$ in Eq.~(\ref{C5}) are determined by
equation of the eigenvalue problem
\begin{eqnarray}
[\hat{\cal L}_n(r) &-& \gamma_m] \, V_{m,n}(r) = 0  \;, \label{C7}
\end{eqnarray}
with
\begin{eqnarray*}
\hat{\cal L}_n(r) &=& {d^2 \over d r^2} + {4 \over r} \,  {d \over
dr} + {2 \Lambda\over r} - {\Lambda^2 \over 4} - {n(n+3) \over r^2}
\; .
\end{eqnarray*}
The constant $\tilde A$ in Eq.~(\ref{C5}) is determined from the
conservation law for the total angular momentum $L_\odot \equiv \int
\{ {\bf r} {\bf \times} [(\Omega + \delta\Omega) {\bf \times} {\bf
r}] \} \, \rho_0({\bf r}) \,d{\bf r}$ of the rotating body (e.g.,
the Sun). The functions $Q_{m,2n}(t)$ in Eq.~(\ref{C4}) are
determined in Appendix B. They are given by
Eqs.~(\ref{C15})-(\ref{C17}).

Equation~(\ref{C7}) coincides with the equation for the Kepler
problem in quantum mechanics (the hydrogen atom in a spherically
symmetric potential, see, e.g., \cite{F71}). The solution of
Eq.~(\ref{C7}) is given by
\begin{eqnarray}
V_{m,n}(r) = r^n \, \exp(- r \, \Lambda / m) \, \tilde F(a; \, b;
\, 2 r \, \Lambda / m) \;, \label{C8}
\end{eqnarray}
where $\tilde F(a; \, b; \, y)$ is the confluent hypergeometric
function with $a=n - m + 2$ and $b = 2(n + 2)$. Here we assumed for
simplicity that $L_\rho$ is independent of the radius. The
characteristic spatial scale of the mean fields variations is of the
order of the solar radius $R_\odot$ in the main part of the
convective zone (except for its boundaries). On the other hand, the
fluid density in the solar convective zone changes very strongly (in
6 - 7 orders of magnitude). Therefore, in the main part of the solar
convective zone $r \, \Lambda \gg 1$, and Eq.~(\ref{C8}) for the
eigenfunctions $V_{m,n}(r)$ reduces to
\begin{eqnarray}
V_{m,n}(r) = A \, r^{m-2} \,  \exp(- r \, \Lambda / m) \;,
\label{C9}
\end{eqnarray}
where the eigenvalues $\gamma_m$ are given by
\begin{eqnarray}
\gamma_m = - {\Lambda^2 (m^2 - 4) \over 4 m^2} \,  \;, \label{C10}
\end{eqnarray}
with the integer numbers $m \geq 3$. Here $\gamma_m$ is measured in
units of $\nu_{_{T}} / R_\odot^{2} $. The constant $A$ in
Eq.~(\ref{C9}) is determined from the condition $\int_{r_b}^1 \, r^4
\, V_{m,n}^2(r) \,dr = 1$, where $r_b = R_b / R_\odot$, $\, R_b$ is
the radius of the bottom of the convective zone,  and $r = R /
R_\odot$ is the dimensionless radius measured in units of the solar
radius $R_\odot$. Therefore, the differential rotation caused by the
effect of the turbulent heat flux on the Reynolds stresses in a
turbulent convection is determined by the following equation:
\begin{eqnarray}
{\delta \Omega \over \Omega} &=& {L_\rho \over R_\odot} \,
\biggl[{\tau_0^2 \, g \, F_\ast \over \nu_{_{T}}}\biggr] \,
\sum_{m=3}^{\infty} \, \biggl\{\beta_m \, \biggl[{\rho_0(R_b) \over
\rho_0(R_\odot)}\biggr]^{m-2 \over 2m}
\nonumber\\
&& - \biggl({R \over R_\odot}\biggr)^{m-2} \, \biggl[{\rho_0(R_b)
\over \rho_0(R)}\biggr]^{m-2 \over 2m} \, [f_{1,m}(\omega)
\nonumber\\
&& + f_{2,m}(\omega) \, \cos^2 \theta] \biggr\} \;, \label{C18}
\end{eqnarray}
where
\begin{eqnarray*}
f_{1,m}(\omega) &=& K(m) \, \biggl[\tilde \Phi_1(\omega) - 3
K_\ast(m) \, {L_\rho \over R_\odot} \, \tilde \Phi_2(\omega)
\biggr] \;,
\\
f_{2,m}(\omega) &=& K(m) \, \Phi_2(\omega) \biggl[1 - 13 K_\ast(m)
\, {L_\rho \over R_\odot} \biggr] \;,
\\
\beta_m &=& {L_\rho \over R_\odot} \, \biggl({10 \, m \over
m-2}\biggr) \, \biggl[{f_1 + f_2 / 5 \over 1 - (R_ b /
R_\odot)^5}\biggr]  \;,
\end{eqnarray*}
$K(m) = 8 m^2 / [3 (m^2 - 4) (m+2)]$, $\, \tilde \Phi_1(\omega) =
\Phi_1(\omega) + (13/10) \Phi_2(\omega)$ and $\, \tilde
\Phi_2(\omega) = \Phi_1(\omega) + 5 \Phi_2(\omega)$, the parameter
$K_\ast(m)$ is given by Eq.~(\ref{B33}) in Appendix B. In the next
section we use Eq.~(\ref{C18}) in order to estimate the solar
differential rotation.

\section{Discussion}

The effect of the turbulent heat flux on the Reynolds stresses in a
turbulent convection may cause the differential rotation comparable
with the typical differential rotation of the Sun. Indeed, let us
use estimates of governing parameters taken from models of the solar
convective zone, e.g., \cite{S74,BT66}. More modern treatments make
little difference to these estimates. In particular, at depth of the
convective zone, $ H \sim 10^{10}$ cm measured from the top (i.e.,
at $R = 0.85 R_\odot$), the parameters are: the maximum scale of
turbulent motions $l_0 \sim 5.5 \times 10^9$ cm; the characteristic
turbulent velocity in the maximum scale of turbulent motions $u_0
\sim 5.4 \times 10^3$ cm s$^{-1}$; the turbulent viscosity
$\nu_{_{T}} \sim 10^{13} $ cm$^2$ s$^{-1}$; the fluid density
$\rho_0 \sim 7.6 \times 10^{-2}$ g cm$^{-3} $ and the fluid density
stratification length $ L_\rho \sim 10^{10}$ cm. Thus,
Eq.~(\ref{C18}) yields the following estimates for the solar
differential rotation
\begin{eqnarray}
\left|{\partial \over \partial r} \delta \Omega \right|&\approx& 200
\, (1 + 0.4 \, \cos^2 \theta) \, {nHz \over R_\odot} \;,
\label{O1}\\
{1 \over r} \, \left|{\partial \over \partial \theta} \delta \Omega
\right|&\approx& 140 \, \sin (2 \theta) \, {nHz \over R_\odot} \; .
\label{O2}
\end{eqnarray}
These estimates are in agreement with the data obtained from surface
observations of the solar angular velocity \cite{HH70,SH84} and from
helioseismology \cite{DH86,DG89,T90,SAB98}. Therefore, the effect of
the turbulent heat flux on the Reynolds stresses in a turbulent
convection is crucial for the formation of the differential rotation
and should be taken into account in theories of the differential
rotations of the Sun and solar-like stars.

The mechanism of the differential rotation due to the effect of the
turbulent heat flux on the Reynolds stresses, is related to the
counter-rotation turbulent heat flux in turbulent convection. This
flux reads:
\begin{eqnarray}
{\bf F}^{\rm CR} &=& - {3 \, F_\ast \over 8 \, \omega} \, \biggl[2\,
\biggl({\arctan \, \omega \over \omega} - 1 \biggr)
\nonumber\\
&& + \; \;  \ln (1 + \omega^{2})\biggr] \, \sin \,\theta \, {\bf
e}_\varphi \;, \label{RB30}
\end{eqnarray}
[see Eq.~(\ref{B30})]. Therefore, the entropy fluctuations correlate
with the toroidal component of the velocity. The entropy
fluctuations result in fluctuations of the buoyancy force, that
increases fluctuations of the poloidal components of the velocity
(which are correlated with the fluctuations of the toroidal
component of the velocity). These produce the off-diagonal
components of the Reynolds stress tensor which cause the formation
of the differential rotation.

\begin{acknowledgments}
This work was initiated during our visit to the Isaac Newton
Institute for Mathematical Sciences (Cambridge) in the framework of
the programme "Magnetohydrodynamics of Stellar Interiors".
\end{acknowledgments}

\appendix

\section{The Reynolds Stresses in rotating turbulent convection}

We use a mean field approach whereby the velocity, pressure and
entropy are separated into the mean and fluctuating parts, where the
equations in the new variables for fluctuations of velocity ${\bf v}
= \sqrt{\rho_0} \, {\bf u}' $ and entropy $s = \sqrt{\rho_0} \, s'$
follow from Eqs.~(\ref{A1}) and (\ref{A2}):
\begin{eqnarray}
{1 \over \sqrt{\rho_0}} {\partial {\bf v}({\bf x},t) \over
\partial t} &=& - \bec{\nabla} \biggl({p' \over \rho_0}\biggr)
+ {1 \over \sqrt{\rho_0}} \biggl[2 {\bf v} \times {\bf \Omega} -
({\bf v} \cdot \bec{\nabla}) {\bf U}
\nonumber \\
&-& G^U \, {\bf v} - {\bf g} \, s \biggr] + {\bf F}_M + {\bf v}^N
\;,
\label{M1} \\
{\partial s({\bf x},t) \over \partial t} &=& - {\Omega_{b}^{2} \over
g} ({\bf v} \cdot {\bf e}) - G^U \, s + s^N \;, \label{M2}
\end{eqnarray}
where $G^U = (1 / 2) \, {\rm div} \, {\bf U} + {\bf U} {\bf \cdot}
\bec{\nabla}$, $\,  {\bf v}^{N} $ and $ s^{N} $ are the nonlinear
terms which include the molecular viscous and dissipative terms,
$p'$ are fluctuations of fluid pressure. The fluid velocity
fluctuations ${\bf v}$ satisfy the equation $\bec{\nabla} \cdot {\bf
v} = (1 / 2 \, L_\rho) ({\bf v} \cdot {\bf e})$, where
$(\bec{\nabla} \rho_0) / \rho_0 = - L_\rho^{-1} \, {\bf e} .$
Equations~(\ref{M1}) and (\ref{M2}) are written in the anelastic
approximation. The variables with the subscript $ "0" $ correspond
to the hydrostatic nearly isentropic basic reference state. The
turbulent convection is regarded as a small deviation from a
well-mixed adiabatic reference state.

Let us derive equations for the second-order moments. For this
purpose we rewrite the momentum equation and the entropy equation in
a Fourier space. In particular,
\begin{eqnarray}
{dv_i({\bf k}) \over dt} &=& [D_{im}^{\Omega}({\bf k}) +
I_{im}^U({\bf k})] v_m({\bf k}) + g \, e_m \, P_{im}({\bf k}) \,
s({\bf k})
\nonumber\\
&& + v_i^N({\bf k}) \;,
\label{B1}\\
{ds({\bf k}) \over dt} &=& - G^U({\bf k}) \, s({\bf k}) + s^N \; .
\label{B2}
\end{eqnarray}
To derive Eq.~(\ref{B1}) we multiplied the momentum equation written
in ${\bf k}$-space by $ P_{ij}({\bf k}) = \delta_{ij} - k_{ij} $ in
order to exclude the pressure term from the equation of motion. Here
\begin{eqnarray*}
I_{ij}^U({\bf k}) &=& 2 k_{in} \nabla_{j} U_{n} - \nabla_{j} U_{i} -
G^U({\bf k}) \delta_{ij} \;,
\\
G^U({\bf k}) &=& {1 \over 2} \, {\rm div} \, {\bf U} + i ({\bf U}
{\bf \cdot} {\bf k}) \;,
\\
D_{ij}^{\Omega}({\bf k}) &=& 2 \varepsilon_{ijm} \Omega_n k_{mn}
\;,
\end{eqnarray*}
and $ \delta_{ij} $ is the Kronecker tensor, $ k_{ij} = k_i  k_j /
k^2$ and $\varepsilon_{ijk}$ is the Levi-Civita tensor. Using
Eqs.~(\ref{B1})-(\ref{B2}) we derive equations for the following
correlation functions:
\begin{eqnarray*}
f_{ij}({\bf k}) &=& \hat L(v_i; v_j) \;, \; \; F_{i}({\bf k}) =
\hat L(s; v_i) \;,
\\
\Theta_{i}({\bf k}) &=& \hat L(s; s) \;,
\end{eqnarray*}
where
\begin{eqnarray*}
\hat L(a; c) = \int \langle a(t,{\bf k} + {\bf  K} / 2) c(t,-{\bf
k} + {\bf  K} / 2) \rangle
\\
\times \exp{(i {\bf K} {\bf \cdot} {\bf R}) } \,d {\bf  K} \;,
\end{eqnarray*}
$ {\bf R} $ and $ {\bf K} $ correspond to the large scales, and $
{\bf r} $ and $ {\bf k} $ to the  small ones. Hereafter we omitted
argument $t$ and ${\bf R}$ in the correlation functions. The
equations for these correlation functions are given by
\begin{eqnarray}
{\partial f_{ij}({\bf k}) \over \partial t} &=& (I_{ijmn}^U +
D_{ijmn}^{\Omega}) f_{mn} + M_{ij}^F + \hat{\cal N} \tilde f_{ij} ,
\label{B3} \\
{\partial F_{i}({\bf k}) \over \partial t} &=& (I_{im}^U +
D_{im}^{\Omega}) F_{m} + g e_m P_{im}(k) \Theta({\bf k}) + \hat{\cal
N} \tilde F_{i} \;,
\nonumber\\
\label{B4} \\
{\partial \Theta({\bf k}) \over \partial t} &=& - {\rm div} \, [{\bf
U} \, \Theta({\bf k})] + \hat{\cal N} \Theta \;, \label{B5}
\end{eqnarray}
where
\begin{eqnarray}
I_{ijmn}^U &=& I^U_{im}({\bf k}_1) \, \delta_{jn} + I^U_{jm}({\bf
k}_2) \, \delta_{in}
\nonumber\\
&=& \biggl[2 k_{iq} \delta_{mp} \delta_{jn} + 2 k_{jq} \delta_{im}
\delta_{pn} - \delta_{im} \delta_{jq} \delta_{np}
\nonumber\\
&& \; - \delta_{iq} \delta_{jn} \delta_{mp} + \delta_{im}
\delta_{jn} k_{q} {\partial \over \partial k_{p}} \biggr] \nabla_{p}
U_{q}
\nonumber\\
&& \; - \delta_{im} \delta_{jn} \, [{\rm div} \, {\bf U} + {\bf U}
{\bf \cdot} \bec{\nabla}]\;,
\label{B18}\\
D_{ijmn}^{\Omega} &=& D_{im}^{\Omega}({\bf k}_1) \, \delta_{jn} +
D_{jm}^{\Omega}({\bf k}_2) \, \delta_{in}
\nonumber\\
&=& 2 \, \Omega_q k_{pq} (\varepsilon_{imp} \, \delta_{jn} +
\varepsilon_{jmp} \, \delta_{in}) \;,
\label{B19}\\
M_{ij}^F &=& g e_m [P_{im}({\bf k}) F_{j}({\bf k}) + P_{jm}({\bf k})
F_{i}(-{\bf k})], \label{B20}
\end{eqnarray}
and ${\bf k}_1 = {\bf k} + {\bf K} / 2$, $\, {\bf k}_2 = -{\bf k} +
{\bf K} / 2 $. Note that the correlation functions $f_{ij}$, $\,
F_{i}$ and $\Theta$ are proportional to the fluid density
$\rho_0({\bf R})$. Here $\hat{\cal N}\tilde f_{ij}$, $\, \hat{\cal
N}\tilde F_{i}$ and $\hat{\cal N}\Theta$ are the terms which are
related to the third-order moments appearing due to the nonlinear
terms. In particular,
\begin{eqnarray*}
\hat{\cal N}\tilde f_{ij} &=& \langle P_{im}({\bf k}_1)
v^{N}_{m}({\bf k}_1) v_j({\bf k}_2) \rangle
\\
&& + \langle v_i({\bf k}_1) P_{jm}({\bf k}_2) v^{N}_{m}({\bf k}_2)
\rangle \;,
\\
\hat{\cal N}\tilde F_{i} &=& \langle s^{N}({\bf k}_1) u_j({\bf k}_2)
\rangle + \langle s({\bf k}_1) P_{im}({\bf k}_2) v^{N}_{m}({\bf
k}_2) \rangle \;,
\\
\hat{\cal N}\Theta &=& \langle s^{N}({\bf k}_1) s({\bf k}_2) \rangle
+ \langle s({\bf k}_1) s^{N}({\bf k}_2) \rangle  \; .
\end{eqnarray*}
When div $\, {\bf U} = 0$, Eq.~(\ref{B18}) coincides with that
derived in \cite{EKRZ02}.

The equations for the second-order moments contain high-order
moments and a closure problem arises (see, e.g.,
\cite{O70,MY75,Mc90}). We apply the spectral $\tau$ approximation
(or the third-order closure procedure, see, e.g.,
\cite{O70,PFL76,KRR90,RK04,BK04,BS05}). The spectral $\tau$
approximation postulates that the deviations of the
third-order-moment terms, $\hat{\cal N}f_{ij}({\bf k})$, from the
contributions to these terms afforded by the background turbulent
convection, $\hat{\cal N}f_{ij}^{(0)}({\bf k})$, are expressed
through the similar deviations of the second moments, $f_{ij}({\bf
k}) - f_{ij}^{(0)}({\bf k})$, i.e.,
\begin{eqnarray}
\hat{\cal N}f_{ij}({\bf k}) - \hat{\cal N}f_{ij}^{(0)}({\bf k}) = -
{f_{ij}({\bf k}) - f_{ij}^{(0)}({\bf k}) \over \tau(k)} \;,
\label{B6}
\end{eqnarray}
and similarly for other tensors, where $\hat{\cal N}f_{ij} =
\hat{\cal N}\tilde f_{ij} + M_{ij}^F(F^{\Omega=0})$ and $\hat{\cal
N} F_{i} = \hat{\cal N}\tilde F_{i} + g e_n P_{in}(k)
\Theta^{\Omega=0} $, the superscript $ (0) $ corresponds to the
background turbulent convection (i.e., a nonrotating turbulent
convection with $ \nabla_{i} \bar U_{j} = 0)$, $\, \tau (k) $ is the
characteristic relaxation time of the statistical moments. The
quantities $F^{\Omega=0}$ and $\Theta^{\Omega=0}$ are for a
nonrotating turbulent convection with nonzero spatial derivatives of
the mean velocity. Note that we applied the $ \tau $-approximation
(\ref{B6}) only to study the deviations from the background
turbulent convection which are caused by the spatial derivatives of
the mean velocity and a nonzero rotation. The background turbulent
convection is assumed to be known (see below).

Solution of Eqs.~(\ref{B3})-(\ref{B5}) after applying the spectral
$\tau$ approximation reads
\begin{eqnarray}
f_{ij}({\bf k}) &=& f^{(0)}_{ij}({\bf k}) + f^{F}_{ij}({\bf k}) +
f^{U}_{ij}({\bf k}) \;,
\label{B7}\\
F_i({\bf k}) &=& \tilde D_{im}^{-1}({\bf \Omega}) F^{(0)}_{m}({\bf
k}) \;,
\label{B8}\\
\Theta({\bf k}) &=& [1 - \tau ({\bf U} {\bf \cdot} \bec{\nabla})]
\Theta^{(0)}({\bf k})  \;,
\label{B9}
\end{eqnarray}
where
\begin{eqnarray}
f^{U}_{ij}({\bf k}) &=& \tau \tilde D_{ijmn}^{-1}({\bf \Omega})
I_{mnpq}^U [f^{(0)}_{pq} + f^{F}_{pq}]  \;,
\label{B11}\\
f^{F}_{ij}({\bf k}) &=& \tau \tilde D_{ijmn}^{-1}({\bf \Omega})
\tilde M_{mn}^F({\bf k}) \;,
\label{B10}\\
\tilde M_{ij}^F({\bf k}) &=& g e_m \{P_{im}({\bf k}) [F_{j}({\bf k})
-F_{j}^{\Omega=0}({\bf k})]
\nonumber\\
&& + P_{jm}({\bf k}) [F_{i}(-{\bf k}) - F_{i}^{\Omega=0}(-{\bf
k})]\} .
\label{TB10}
\end{eqnarray}
Here $\tilde D_{ij}^{-1}({\bf \Omega})$ is the inverse of $\tilde
D_{ij}({\bf \Omega}) = \delta_{ij} - \tau D_{ij}^{\Omega}$ and
$\tilde D_{ijmn}^{-1}({\bf \Omega})$ is the inverse of $\tilde
D_{ijmn}({\bf \Omega}) = \delta_{im} \delta_{jn} - \tau \,
D_{ijmn}^{\Omega}$, and
\begin{eqnarray}
&& \tilde D_{ij}^{-1}({\bf \Omega}) = \chi(\psi) \, (\delta_{ij} +
\psi \, \varepsilon_{ijm} \, \hat k_m + \psi^2 \, k_{ij}) \;,
\label{B12}\\
&& \tilde D_{ijmn}^{-1}({\bf \Omega}) = {1 \over 2} [B_1 \,
\delta_{im} \delta_{jn} + B_2 \, k_{ijmn} + B_3 \,
(\varepsilon_{ipm} \delta_{jn}
\nonumber\\
&& + \varepsilon_{jpn} \delta_{im}) \hat k_p + B_4 \, (\delta_{im}
k_{jn} + \delta_{jn} k_{im})
\nonumber\\
&& + B_5 \, \varepsilon_{ipm} \varepsilon_{jqn} k_{pq} + B_6 \,
(\varepsilon_{ipm} k_{jpn} + \varepsilon_{jpn} k_{ipm}) ] \;,
\label{B14}
\end{eqnarray}
and $\hat k_i = k_i / k$, $\, \chi(\psi) = 1 / (1 + \psi^2) $, $\,
\psi = 2 \tau(k) \, ({\bf k} \cdot {\bf \Omega}) / k $, $\, B_1 = 1
+ \chi(2 \psi) ,$ $\, B_2 = B_1 + 2 - 4 \chi(\psi) ,$ $\, B_3 = 2
\psi \, \chi(2 \psi) ,$ $\, B_4 = 2 \chi(\psi) - B_1 ,$ $\, B_5 = 2
- B_1 $ and $B_6 = 2 \psi \, [\chi(\psi) - \chi(2 \psi)] $. For
derivation of Eqs.~(\ref{B7}) and~(\ref{B11})-(\ref{B10}) we used a
procedure described in Appendix B in \cite{EGKR05}.

For the integration in ${\bf k}$-space of the second moments
$f_{ij}({\bf k})$, $\, F_{i}({\bf k})$ and $\Theta({\bf k}) $ we
have to specify a model for the background turbulent convection
(i.e., a nonrotating turbulent convection with $ \nabla_{i} \bar
U_{j} = 0)$. Here we used the following model of the background
turbulent convection:
\begin{eqnarray}
f_{ij}^{(0)}({\bf k}) &=& \rho_0 \, \langle ({\bf u}')^2 \rangle
P_{ij}({\bf k}) W(k) \;,
\label{B15} \\
F^{(0)}_{i}({\bf k}) &=& 3 \, \rho_0 \, \langle s' \, u'_z \rangle
\, e_{m} P_{im}({\bf k}) \, W(k) \;,
\label{B16} \\
\Theta^{(0)}({\bf k}) &=& 2 \, \rho_0 \, \langle (s')^2 \rangle \,
W(k) \;,
\label{B17}
\end{eqnarray}
where $W(k) = E(k) / 8 \pi k^{2} ,$ $\, \tau(k) = 2 \tau_0 \bar
\tau(k) ,$ $ \, E(k) = - d \bar \tau(k) / dk ,$ $ \, \bar \tau(k) =
(k / k_{0})^{1-q} ,$ $ \, 1 < q < 3 $  is the exponent of the
kinetic energy spectrum $ (q = 5/3 $ for Kolmogorov spectrum), $
k_{0} = 1 / l_{0} ,$ and $ l_{0} $ is the maximum scale of turbulent
motions, $ \tau_0 = l_{0} / u_{0} $ and $ u_{0} $ is the
characteristic turbulent velocity in the scale $ l_{0} .$ Motion in
the background turbulent convection is assumed to be non-helical.

Equations~(\ref{B8}), (\ref{B10}) and (\ref{TB10}) can be rewritten
in the form
\begin{eqnarray}
F_{i}({\bf k}) &=& \rho_0 \, F_\ast \, {3 \,  W(k) \over 1 +
\psi^2} \, [e_{m} P_{im}({\bf k}) + \psi \, ({\bf e} {\bf \times}
{\bf k})_{i}] \;,
\nonumber\\
\label{B21}\\
f^{F}_{ij}({\bf k}) &=& \rho_0 \, \tau({\bf k}) \, g \,   F_\ast \,
{3 \psi \, W(k) \over 2(1 + \psi^2)} \, [B_1 \, (M_{ij}^{(a)}
\nonumber\\
&& - 2 \psi M_{ij}^{(c)}) + 2 \psi (B_1 - 2) \, M_{ij}^{(b)}] \;,
\label{B23}\\
\tilde M_{ij}^F({\bf k}) &=& \rho_0 \, \tau({\bf k}) \, g \, F_\ast
\, {3 \psi \, W(k) \over 1 + \psi^2} \, [M_{ij}^{(a)} - 2 \psi
M_{ij}^{(b)}] \;,
\nonumber\\
\label{B22}
\end{eqnarray}
where $F_\ast = \langle s' \, u'_z \rangle$,
\begin{eqnarray*}
M_{ij}^{(a)} &=& ({\bf e} {\bf \times} {\bf k})_{i} \, e_{m} \,
P_{jm}({\bf k}) + ({\bf e} {\bf \times} {\bf k})_{j} \, e_{m} \,
P_{im}({\bf k}) \;,
\\
M_{ij}^{(b)} &=& e_m \, e_n \, P_{im}({\bf k}) \, P_{jn}({\bf k})
\;,
\\
M_{ij}^{(c)} &=& ({\bf e} {\bf \times} {\bf k})_{i} \, ({\bf e} {\bf
\times} {\bf k})_{j} \;,
\end{eqnarray*}
and we used the identities:
\begin{eqnarray*}
\tilde D_{ijmn}^{-1} \, M_{ij}^{(a)} &=& {1 \over 2} [( B_1 - B_5)\,
M_{ij}^{(a)}
\\
&& + 2 B_3 \, (M_{ij}^{(b)} - M_{ij}^{(c)})] \;,
\\
\tilde D_{ijmn}^{-1} \, M_{ij}^{(b)} &=& {1 \over 2} [B_1 \,
M_{ij}^{(b)}  - B_3 \, M_{ij}^{(a)} + B_5 \, M_{ij}^{(c)}] \; .
\end{eqnarray*}

In order to integrate over the angles in $ {\bf k} $-space we used
the following identities:
\begin{eqnarray}
&& \bar J_{ij}(a) = \int {k_{ij} \sin \theta \over 1 + a \cos^{2}
\theta} \,d\theta \,d\varphi =  \bar A_{1} \delta_{ij} + \bar A_{2}
\, \omega_{ij} \;,
\nonumber\\
\label{A7}\\
&& \bar J_{ijmn}(a) = \int {k_{ijmn} \sin \theta \over 1 + a
\cos^{2} \theta} \,d\theta \,d\varphi = \bar C_{1} (\delta_{ij}
\delta_{mn} + \delta_{im} \delta_{jn}
\nonumber\\
&& + \delta_{in} \delta_{jm}) + \bar C_{2} \, \omega_{ijmn} + \bar
C_{3} (\delta_{ij} \omega_{mn} + \delta_{im} \omega_{jn} +
\delta_{in} \omega_{jm}
\nonumber\\
&& + \delta_{jm} \omega_{in} + \delta_{jn} \omega_{im} + \delta_{mn}
\omega_{ij}) \;, \label{A8}
\end{eqnarray}
(see for details, \cite{EGKR05,KR03}), where $\omega_{ij} =
\Omega_{i} \, \Omega_{j} / \Omega^2$, $\, \omega_{ijmn} =
\omega_{ij} \, \omega_{mn}$,
\begin{eqnarray*}
\bar A_{1}(a) &=& {2 \pi \over a} \biggl[(a + 1) {\arctan (\sqrt{a})
\over \sqrt{a}} - 1 \biggr] \;,
\\
\bar A_{2}(a) &=& - {2 \pi \over a} \biggl[(a + 3) {\arctan
(\sqrt{a}) \over \sqrt{a}} - 3 \biggr] \;,
\\
\bar C_{1}(a) &=& {\pi \over 2a^{2}} \biggl[(a + 1)^{2} {\arctan
(\sqrt{a}) \over \sqrt{a}} - {5 a \over 3} - 1 \biggr] \;,
\\
\bar C_{2}(a) &=& \bar A_{2}(a) - 7 \, \bar A_{1}(a) + 35 \, \bar
C_{1}(a) \;,
\\
\bar C_{3}(a) &=& \bar A_{1}(a) - 5 \bar C_{1}(a)  \; .
\end{eqnarray*}

Equation~(\ref{B23}) after the integration in ${\bf k}$-space allows
us to determine the contributions of the turbulent heat flux to the
Reynolds stress tensor in turbulent convection. In particular, the
components $\sigma_{r \varphi}^F = - e_j^\varphi \, e_i^r \,
f^{F}_{ij}$ and $\sigma_{\theta \varphi}^F = - e_j^\varphi \,
e_i^\theta \, f^{F}_{ij}$ are given by Eqs.~(\ref{B24}) and
~(\ref{B25}), respectively, where
\begin{eqnarray}
\Phi_1(\omega) &=& 2 \Psi_1(\omega) + \Psi_2(\omega/2) \;,
\label{B31}\\
\Phi_2(\omega) &=& 2 \Psi_2(\omega) + \Psi_2(\omega/2) \;,
\label{B32}
\end{eqnarray}
\begin{eqnarray*} \Psi_1(\omega) &=& - {6 \over \omega^{4}}
\biggl[{\arctan \, \omega \over \omega} (1 + \omega^{2}) - {8
\omega^{2}\over 3} -1
\\
&& + 2 \, \omega \, S(\omega) \biggr] \;,
\\
\Psi_2(\omega) &=& {6 \over \omega^{4}} \biggl[5 \, {\arctan \,
\omega \over \omega} (1 + \omega^{2}) + {8 \omega^{2}\over 3} - 5
\\
&& - 6 \, \omega \, S(\omega) \biggr] \;,
\end{eqnarray*}
$\omega = 8 \tau_0 \Omega $ and $S(\omega) = \int_{0}^{\omega}
[\arctan \, y / y] \,d y$. Here we took into account that
$\Psi_1(\omega) = A_1^{(2)}(\omega) + A_2^{(2)}(\omega) -
C_1^{(2)}(\omega) - C_3^{(2)}(\omega)$ and $\Psi_2(\omega) = -
C_2^{(2)}(\omega) - 3 C_3^{(2)}(\omega)$, where the functions
$A_k^{(p)}(\omega)$ and $C_k^{(p)}(\omega)$ are given by
\begin{eqnarray*}
A_{k}^{(p)}(\omega) &=& (6 / \pi \omega^{p+1}) \int_{0}^{\omega}
y^{p} \bar A_{k}(y^{2}) \,d y \;,
\\
C_{k}^{(p)}(\omega) &=& (6 / \pi \omega^{p+1}) \int_{0}^{\omega}
y^{p} \bar C_{k}(y^{2}) \,d y \;,
\end{eqnarray*}
(see for details, \cite{EGKR05,KR03}). For derivation of
Eqs.~(\ref{B24}) and ~(\ref{B25}) we used the following identities:
\begin{eqnarray*}
&& e_j^\varphi \, e_i^\theta \, M_{ij}^{(a)} = k_{ijm}  \,
(e_i^\varphi \, e_j^\varphi - e_i^\theta \, e_j^\theta) e_m^r  \;,
\\
&& e_j^\varphi \, e_i^\theta \, M_{ij}^{(b)} = k_{ijmn}  \,
e_i^\varphi \, e_j^\theta \, e_m^r \, e_n^r \;,
\\
&& e_j^\varphi \, e_i^\theta \, M_{ij}^{(c)} = - k_{ij}  \,
e_j^\varphi \, e_i^\theta \;,
\\
&& (e_i^\varphi \, e_j^\varphi - e_i^\theta \, e_j^\theta)  \, e_m^r
\hat \omega_n  \, \bar J_{ijmn} = (\bar C_2 + 3 \bar C_3) \, \sin^2
\theta \, \cos \theta ,
\\
&& \varepsilon_{pqj} \, e_m^r \, e_n^r \, e_q^r \, \hat \omega_i
\bar J_{ijmn} = ({\bf e} {\bf \times} \bec{\hat \omega})_p [\bar C_1
+ \bar C_3
\nonumber\\
&& + \cos^2 \theta \, (\bar C_2 + 3 \bar C_3)] \; .
\end{eqnarray*}

Equation for the turbulent heat flux follows from Eq.~(\ref{B21})
after integration in ${\bf k}$-space:
\begin{eqnarray}
{\bf F} &=& {F_\ast \over 16} \biggl(- x \, [A_1^{(1)}(x) +
A_2^{(1)}(x)] \, \sin \theta \, {\bf e}^\varphi + [2 A_1^{(0)}(x)
\nonumber\\
&& + A_2^{(0)}(x) \, \sin^2 \theta] \, {\bf e}^r + {1 \over 2} \,
A_2^{(0)}(x) \sin (2 \theta)  \, {\bf e}^\theta \biggr)_{x=\omega/2}
\;,
\nonumber\\
\label{B30}
\end{eqnarray}
where ${\bf e}^r$,  $\, {\bf e}^\theta$ and $\, {\bf e}^\varphi$ are
the unit vectors along the radial, meridional and toroidal
directions, respectively. The first term in Eq.~(\ref{B30})
describes the counter-rotation turbulent heat flux, which is given
by ~(\ref{RB30}).

\bigskip

\section{The functions $Q_{m,2n}(t)$}

Equation for the functions $Q_{m,2n}(t)$ follows from
Eqs.~(\ref{C4})-(\ref{C7}). In particular,
\begin{eqnarray}
{d Q_{m,0}(t) \over d t} &=& - |\gamma_m| \, Q_{m,0} + I_0 \,
\biggl[{R_\odot \over L_\rho} \, K_1 (m) \, \Phi_3(\omega)
\nonumber\\
&& + K_2 (m) \, \Phi_4(\omega) \biggr] \;,
\label{C11}\\
{d Q_{m,2}(t) \over d t} &=& - |\gamma_m| \, Q_{m,2} - {2 \over
15} \, I_0 \, \Phi_2(\omega) \, \biggl[{R_\odot \over L_\rho} \,
K_1 (m) \,
\nonumber\\
&& - 13 K_2 (m) \biggr] \;,
\label{C12}\\
{d Q_{m,2n}(t) \over d t} &=& - |\gamma_m| \, Q_{m,2n} \;,
\label{C14}
\end{eqnarray}
with the integer numbers $n>1$ in Eqs.~(\ref{C14}), where
$\Phi_3(\omega) = - \Phi_1(\omega) - (3/2) \Phi_2(\omega)$, $\,
\Phi_4(\omega) = 3 \Phi_1(\omega) + (35/2) \Phi_2(\omega)$, $\, K_1
(m) = \int_{r_b}^1 \, r^3 \, \sqrt{\rho_0(r)} \, V_{m,2n}(r) \,dr $
and $K_2 (m) = \int_{r_b}^1 \, r^2 \, \sqrt{\rho_0(r)} \,
V_{m,2n}(r) \,dr $. A steady state solution of
Eqs.~(\ref{C11})-(\ref{C14}) is given by
\begin{eqnarray}
Q_{m,0}(t) &=& {I_0 \over |\gamma_m|} \, \biggl[{R_\odot \over
L_\rho} \, K_1 (m) \, \Phi_3(\omega) + K_2 (m) \, \Phi_4(\omega)
\biggr] \;,
\nonumber\\
\label{C15}\\
Q_{m,2}(t) &=& - {2 \, I_0 \over 15 \, |\gamma_m|}  \,
\Phi_2(\omega) \, \biggl[{R_\odot \over L_\rho} \, K_1 (m) - 13
K_2 (m) \biggr] \;,
\nonumber\\
\label{C16}\\
Q_{m,2n}(t) &=& 0 \;, \label{C17}
\end{eqnarray}
where $ n > 1$ in Eqs.~(\ref{C17}). The ratio $K_\ast(m) = K_2 (m) /
K_1 (m)$ is given by
\begin{eqnarray}
K_\ast(m) = {r_b^2 \, E_m(r_0) - 1 \over r_b^3 \, E_m(r_0) - 1} \;,
\label{B33}
\end{eqnarray}
where
\begin{eqnarray*}
E_m(r_0) = \exp \biggl( {(2+m)(1 - r_b) R_\odot \over 2m L_\rho}
\biggr) \;,
\end{eqnarray*}
and $r_b = R_b / R_\odot$.

The function $C_n^{3/2}(X)$ entering in Eq.~(\ref{C5}) has the
following properties:
\begin{eqnarray}
&& \int_{-1}^1 \, (1-X^2) \, C_n^{3/2}(X) \, C_l^{3/2}(X) \,dX
\nonumber\\
&& = {(n+1)(n+2) \over n+3/2} \, \delta_{nl} \;, \label{B34}
\end{eqnarray}
and $C_0^{3/2}(X) = 1$, $\, C_2^{3/2}(X) = (3/2) (5X^2 - 1)$. Note
that due to the condition~(\ref{B34}), the function $C_0^{3/2}(X)$
 only contributes to the total angular momentum of the Sun.

\end{document}